\documentclass[%
 aps,
 prapplied,
 amsmath,amssymb,
 reprint,%
 superscriptaddress,
]{revtex4-2}

\usepackage{graphicx}
\usepackage{bm}

\usepackage[utf8]{inputenc}
\usepackage[T1]{fontenc}
\usepackage{etoolbox}
\usepackage{siunitx}
\usepackage[version=4]{mhchem} 
\usepackage{hyperref}
\usepackage{placeins} 
\usepackage{capt-of} 

\usepackage{amsmath}
\usepackage{circuitikz}
\ctikzset{bipoles/thickness=3}
\usepackage{pgfplots}
\pgfplotsset{compat=1.18}

\begin{document}

\title[]{Photo-induced currents and short-term memory for reservoir computing in a ferroelectric semiconductor
}
\author{Yan Meng Chong}
 \thanks{These authors contributed equally to this work.}
 \affiliation{Department of Materials Science and Engineering, Norwegian University of Science and Technology (NTNU), Trondheim, Norway}
\author{Atreya Majumdar}
 \thanks{These authors contributed equally to this work.}
 \affiliation{Faculty of Physics and Center for Nanointegration Duisburg-Essen (CENIDE), University of Duisburg-Essen, Duisburg, Germany}
\author{Manuel Zahn}
    \affiliation{Department of Materials Science and Engineering, Norwegian University of Science and Technology (NTNU), Trondheim, Norway}
    \affiliation{Experimental Physics V, Center for Electronic Correlations and Magnetism, University of Augsburg, Augsburg, Germany}
\author{Ingvild Hansen}
 \affiliation{Department of Materials Science and Engineering, Norwegian University of Science and Technology (NTNU), Trondheim, Norway}
\author{Karin Everschor-Sitte}
 \affiliation{Faculty of Physics and Center for Nanointegration Duisburg-Essen (CENIDE), University of Duisburg-Essen, Duisburg, Germany}
\author{Dennis Meier}
 \email[Contact author:~]{dennis.meier@uni-due.de}
 \affiliation{Department of Materials Science and Engineering, Norwegian University of Science and Technology (NTNU), Trondheim, Norway}
 \affiliation{Faculty of Physics and Center for Nanointegration Duisburg-Essen (CENIDE), University of Duisburg-Essen, Duisburg, Germany}
 \affiliation{Research Center Future Energy Materials and Systems, Research Alliance Ruhr, Bochum, Germany.}

\date{\today}

\begin{abstract}

Physical reservoir computing represents an energy-efficient approach for processing temporal signals by exploiting the intrinsic nonlinear dynamics and fading memory of a physical system. Recently, ferroelectric semiconductors moved into focus as reservoir materials motivated by their versatile electronic responses to external stimuli. Here, we explore the fundamental possibility to recognize time-varying light pulses via photo-induced currents, using the small-band-gap p-type semiconductor ErMnO$_3$ as a model system. Under white-light illumination, ErMnO$_3$ exhibits nonlinearly evolving photo-induced currents and controllable relaxation dynamics that naturally realize the high-dimensional projection and fading memory capabilities required for reservoir computing. The reservoir capability of ErMnO$_3$ is reflected by the improved recognition accuracy of ``Past'' input pulses, which increases from $\sim 33\%$ to $\sim 93\%$ after applying reservoir transformation to the input signal. The results present ferroelectric hexagonal manganites as a promising platform for photo-induced current-based reservoir computing and highlight the potential of light-driven oxide semiconductors for temporal information processing.

\end{abstract}

\maketitle


\begin{figure*}[!t]
\includegraphics{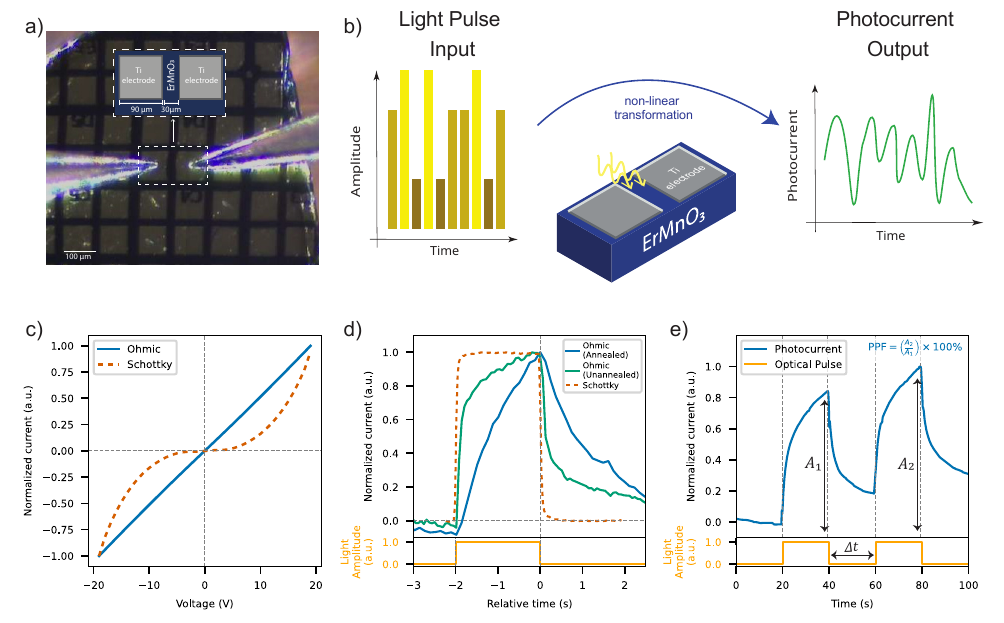}
\caption{\label{fig1}Photo-induced current response of ErMnO$_3$.
(a) Microscopy image of the two-probe measurement on ErMnO$_3$ with laterally deposited Ti electrodes.
(b) Experimental overview illustrating the nonlinear transformation of an input light-pulse sequence into a photo-induced current response, with ErMnO$_3$ acting as the reservoir.
(c) I-V measurements of ErMnO$_3$ forming Ohmic-like and Schottky-like contact. 
(d) Photo-induced current relaxation for three contact types, giving tunable response time. Fastest rise and decay in Schottky contact followed by unannealed Ohmic contact and lastly the reduction annealed Ohmic contact in (5\% H$_2$/N$_2$).
(e) Paired-pulse facilitation (PPF) of the annealed Ohmic sample: the normalized photo-induced current (top) in response to two light pulses separated by $\Delta t$ (bottom). PPF is the ratio of the second to first current-peak amplitude, $\mathrm{PPF}=(A_2/A_1)\times 100\%$. Since $\Delta t$ is shorter than the single-pulse relaxation time, residual current from the first pulse adds to the second, giving $A_2 > A_1$ and $\mathrm{PPF} > 100\%$, which evidences short-term memory.
}
\end{figure*}

\begin{figure*}[!t]
\includegraphics{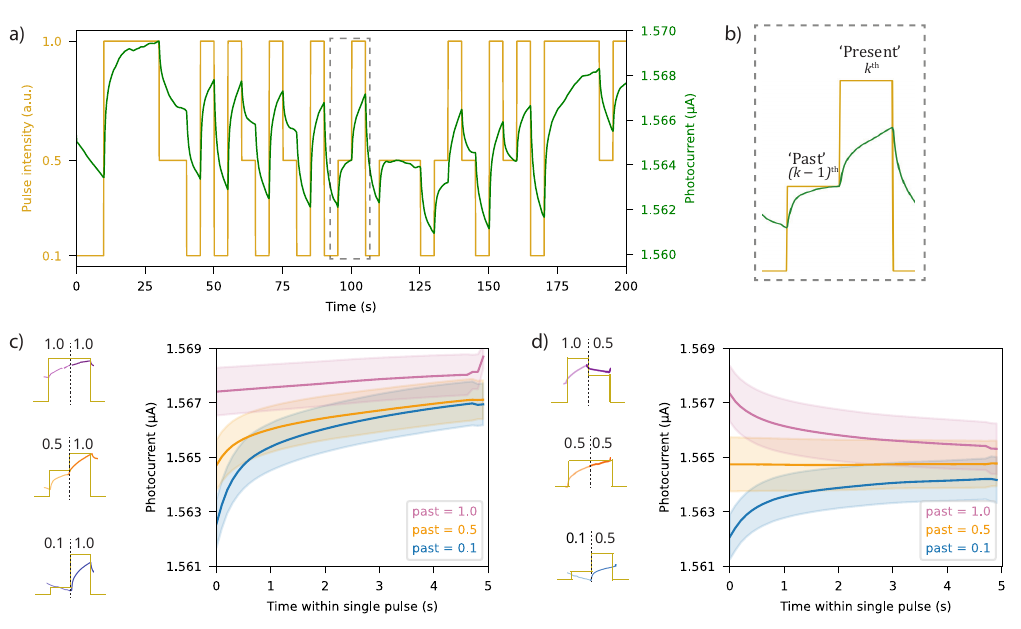}
\caption{\label{fig2}Short-term memory in the photo-induced current of ErMnO$_3$.
(a) Randomly ordered light-pulse sequence with three intensity levels (0.1, 0.5, 1.0) overlaid with the corresponding photo-induced current response of ErMnO$_3$. 
(b) Two-pulse configuration, exemplarily highlighted by the dashed box in (a), the pulses are defined as the ‘Present’ input, $u(k)$ (the $k$-th pulse), and the immediately preceding ‘Past’ input, $u(k-1)$ (pulse $k-1$).
(c) Average photo-induced current traces for all two-pulse configurations with a present pulse of intensity level 1.0. The three possible input pairs are [0.1, 1.0], [0.5, 1.0], and [1.0, 1.0].
(d) Same analysis as in (c), but for present-pulse intensity level 0.5, showing the corresponding three two-pulse configurations.
}
\end{figure*}

\begin{figure*}[!t]
\includegraphics{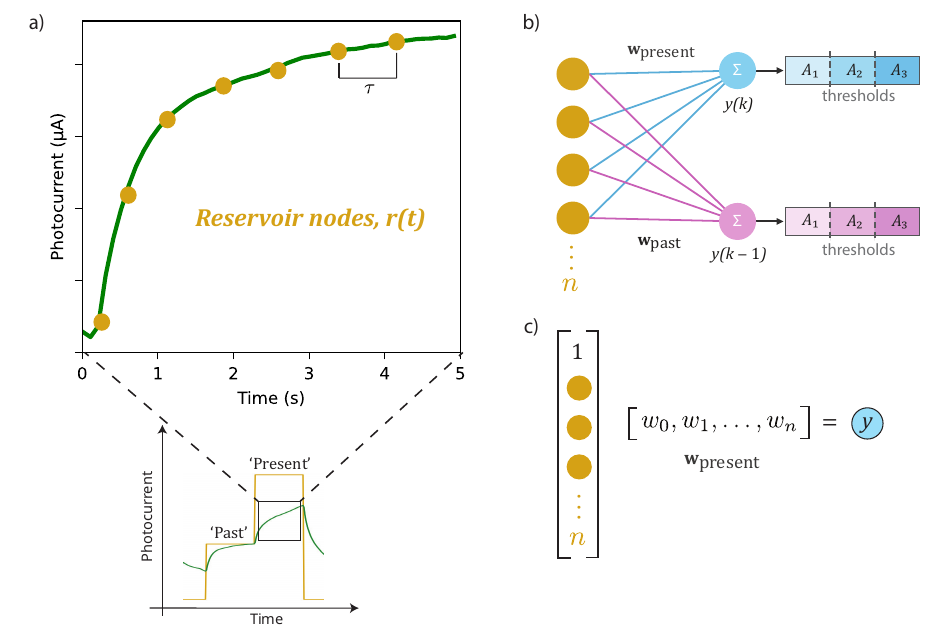}
\caption{\label{fig3}
Reservoir readout for pulse recognition using virtual nodes.
(a) Photo-induced current in response to a \SI{5}{\second} light pulse. Multiple readouts (gold circles), taken at successive time points spaced by the sampling interval $\tau$, serve as the reservoir nodes.
(b)  Linear readout layer receiving the reservoir nodes as input. Two weight vectors, $\mathbf{w}_{\text{present}}$ and $\mathbf{w}_{\text{past}}$, map the reservoir state to the present and past predictions, $y(k)$ and $y(k-1)$. Each prediction is a single least-squares regressed value, thresholded at the midpoints between adjacent intensity levels to assign the pulse to one of the three levels $A_1$, $A_2$, or $A_3$.
(c) Vector representation of a readout, shown here for the present prediction $y(k)$: the output is the inner product of the trained weight vector $\mathbf{w}_{\text{present}}$ with the reservoir state vector. The past prediction $y(k-1)$ is obtained in the same way using $\mathbf{w}_{\text{past}}$. The weight vectors are obtained directly from the closed-form least-squares solution of this regression.
}
\end{figure*}

\begin{figure}[!t]
\includegraphics{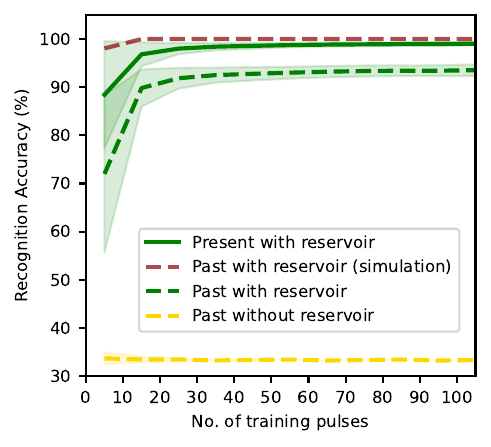}
\caption{\label{fig4}
Recognition accuracy for identifying the present and past light pulses as a function of the number of training pulses. Solid green: accuracy of recognizing the present input using the reservoir. Dashed green: accuracy of recognizing the past input using the reservoir. Dashed yellow: accuracy of recognizing the past input without the reservoir, obtained by training directly on the input intensities. Dashed red: accuracy of recognizing past input with simulation. For each training-set size, the curves show the mean recognition accuracy over 1000 independent random train/test splits, and the shaded bands represent the standard deviation across these splits.}
\end{figure}


\section{\label{sec:level1}Introduction}

 Reservoir computing (RC)~\cite{Lukosevicius2009,Lukosevicius2012} is an energy-efficient computing paradigm that exploits a complex dynamical system to process temporal data, such as speech signals~\cite{Verstraeten2005}, chaotic time series (Mackey-Glass, Lorenz)~\cite{Jaeger2004,Pathak2018}, weather forecasting~\cite{Arcomano2020} and other sequential signals~\cite{Verstraeten2007}. An RC system consists of three key components: an input layer, a reservoir layer, and a linear readout layer as explained in detail elsewhere (see, e.g., Refs.~\onlinecite{Lukosevicius2009,Tanaka2019}). Importantly, training is performed only on the readout layer, which is why the training cost is substantially reduced compared to conventional feed-forward and recurrent neural networks~\cite{Lukosevicius2012}.

 The effectiveness of RC, however, critically relies on the dynamical richness of the reservoir layer. This layer should exhibit four key properties, that is, nonlinearity, high complexity, fading memory, and reproducibility~\cite{Tanaka2019,Jaeger2001}. Notably, these requirements can be fulfilled by physical reservoirs leveraging their response to external stimuli. This concept has been demonstrated across a wide range of platforms, from simple mechanical systems, such as a bucket of water~\cite{Fernando2003}, to technologically relevant implementations including memristors~\cite{Du2017,Moon2019}, spintronic oscillators~\cite{Torrejon2017}, magnetic textures~\cite{Prychynenko2018,Yokouchi2022,Msiska2023,Raab2022} and photonic devices~\cite{Appeltant2011}. Comprehensive overviews of physical reservoir implementations are available in recent reviews (e.g., Refs.~\onlinecite{Tanaka2019,Yan2024,Liang2024,lee2023}). Recently, electric and magnetic topological defects in ferroics, such as skyrmions~\cite{Pinna2020} and domain walls~\cite{EverschorSitte2024}, have been proposed as nanoscale reservoirs, giving additional opportunities for designing physical reservoirs with enhanced complexity and minimal system size. Ferroelectric oxide semiconductors are particularly appealing in this context. Beyond their versatile, electrically and optically tunable responses, they offer the long-term prospect of back-end-of-the-line integration with CMOS, co-locating physical computing elements directly above silicon logic and thereby reducing data-movement overheads. 

Here, we investigate a ferroelectric p-type semiconductor as a physical reservoir by performing a recognition task on temporal light-driven signals. We use ErMnO$_3$, a hexagonal manganite, as our model system. It has been intensively studied for its electric~\cite{Aken2004,Choi2010}, magnetic~\cite{Wood1973}, and multiferroic properties~\cite{Fiebig2002}. Furthermore, ErMnO$_3$ as a semiconductor is well understood in terms of its electronic properties and p-type conduction mechanism~\cite{Sandra2016,Holstad2018}, including versatile functional responses at the local scale~\cite{Jiali2025,Evans2020,Schaab2018,Geng2012,Mundy2017}. Importantly, the material exhibits rich responses to external stimuli such as electric fields, magnetic fields, strain and, most interestingly for this work, to light, enabling generation of sizable photo-induced currents~\cite{Sheng2021}. With a narrow band gap of 1.3–\SI{1.6}{eV}~\cite{Choi2008a,Huang2015}, ErMnO$_3$ efficiently absorbs visible light, enabling the photocarrier excitation which we leverage in this work.

On ErMnO$_3$, we deposit metal electrodes and measure its photo-induced current response. We observe improved recognition accuracy of ``Past'' light pulses with ErMnO$_3$ as the physical reservoir, which we attribute to its nonlinear, time-evolving photo-induced relaxation. We further find that the characteristic timescale of this relaxation (the reservoir's response time) can be tuned: donor doping shifts the work function of ErMnO$_3$ and drives the metal–semiconductor contact from Schottky-like to Ohmic-like, while defect engineering via reducing-atmosphere annealing prolongs the photocurrent relaxation. This tunability modifies the photo-induced current response and, in turn, the reservoir's memory. Our results present this small-band-gap ferroelectric semiconductor as a light-driven physical reservoir for recognition in temporal tasks with adjustable response time.

\section{Results and Discussion}

For our experiments, we prepare two-terminal structures for electrical read-out of light-induced signals as presented in Fig.~\ref{fig1}. For this purpose, a set of titanium electrodes is deposited onto the polished surface of a bulk ErMnO$_3$ crystal (for details of the crystal growth, see Ref.~\cite{Zewu2015}). The electrodes are square-shaped (90 $\times$ \SI{90}{\micro\meter}$^2$), separated by a gap of about \SI{30}{\micro\meter} (Fig.~\ref{fig1}(a)). To test the reservoir capabilities, two neighboring electrodes are selected and a fixed voltage of \SI{2}{V} is applied between them. Photocurrents are then excited using white light (24 programmable NeoPixels RGB LEDs arranged in a circle) to illuminate the gap between the electrodes as schematically illustrated in Fig.~\ref{fig1}(b).

The photo-induced current measured for a single light pulse is presented in Fig.~\ref{fig1}(d), considering three examples, that is, a Schottky-like contact on ErMnO$_3$ and Ohmic-like contacts on as-grown and on N$_2$/H$_2$-annealed ErMnO$_3$. ``Single light pulse'' here refers to continuous illumination of the sample for \SI{2}{s} before the LED is switched off again as shown in the lower panel of Fig.~\ref{fig1}(d). In response, three distinct photocurrent curves are observed depending on the nature of the metal-semiconductor contact. The Schottky-like contact exhibits the shortest rise and decay times, followed by the Ohmic-like contact on as-grown ErMnO$_3$, whereas the Ohmic-like contact on N$_2$/H$_2$-annealed ErMnO$_3$ shows the slowest response. The characteristic rise and decay times range from \SI{20}{ms} to several seconds.

In general, the nature of the metal-semiconductor contact is determined primarily by the relative work functions of the metal, $\phi_\text{m}(\mathrm{Ti})$ and the semiconductor, $\phi_\text{s}(\mathrm{ErMnO_3})$. For p-type ErMnO$_3$, the work function is approximately $\sim 4.36$~eV~\cite{Chen2020} and can be readily tuned through doping and defect engineering~\cite{Didrik2025}. Our as-grown ErMnO$_3$ single crystals form a Schottky-like contact with Ti electrodes because $\phi_\text{m}(\mathrm{Ti}) < \phi_\text{s}(\mathrm{ErMnO_3})$. Corresponding I-V curves are displayed in Fig.~\ref{fig1}(c). Upon donor doping, the work function of ErMnO$_3$ decreases, resulting in $\phi_\text{m}(\mathrm{Ti}) > \phi_\text{s}(\mathrm{ErMnO_3})$, driving the system towards Ohmic-like behavior.

The observed differences in the photocurrent response time in Fig.~\ref{fig1}(d) originate from the distinct carrier-transport mechanisms associated with different types of metal-semiconductor contacts. In the case of a Schottky-like contact, carriers are predominantly generated within the depletion region, where the strong built-in field rapidly separates and sweeps carriers across the junction and recombination at the metal interface proceeds quickly once the illumination is removed. In contrast, in Ohmic systems, photocurrent arises primarily from photogenerated carriers in the bulk, where carrier lifetimes are governed by slower trap-assisted capture and release processes. Annealing the Ohmic-like system in a reducing atmosphere (here, N$_2$ / H$_2$ at 300$^\circ$C for 48 hours) prolongs the photocurrent response time, as oxygen vacancies and additional defects are introduced, creating more trap states for photo-generated carriers~\cite{Guo2014}. The evolution of photocurrents in Fig.~\ref{fig1}(d) can thus be explained based on the nature of the metal-semiconductor contacts, which can be leveraged to achieve the desired response time.

Importantly for this work, the gradual rise of the photocurrent under illumination produces a time-evolving photo-induced current trajectory in response to a single pulse, so that sampling the current at several time points across the response window yields multiple readouts from a single input. This method is analogous to the dimension expansion in delay-based, time-multiplexed reservoir systems introduced in Ref.~\onlinecite{Appeltant2011}, in which a single physical node is unfolded into multiple virtual nodes. The relaxation after the light is turned off provides short-term memory, evidenced in Fig.~\ref{fig1}(e) by a paired-pulse facilitation (PPF) index exceeding \SI{100}{\percent}: the amplitude of the second photo-induced current pulse exceeds the first when the pulse interval $\Delta t$ is shorter than the single-pulse relaxation time.

Together, these features fulfill key requirements of a physical reservoir: the nonlinear transformation of light into photo-induced current, the dimensional expansion provided by the time-evolving response, and the short-term memory encoded in the relaxation dynamics. Next, to perform the pulse recognition task, we select the as-grown Ohmic-ErMnO$_3$ system and illuminate it with a randomly ordered sequence of light pulses of three distinct intensity levels. The three intensity levels are defined as fractions of the maximum light intensity, so that $A_1=0.1,\,A_2=0.5,\,A_3=1.0$ and the level is varied every \SI{5}{s} as shown in Fig.~\ref{fig2}(a). We measure the photocurrent output in response to the light pulse sequence and the overlay of the input (yellow) and output (green) is presented in Fig.~\ref{fig2}(a). Prior to analysis, a slow baseline drift of the measured current, which is uninformative about the immediate pulse history, is removed by background subtraction (see Appendix~\ref{app:baseline}). As expected, the photocurrent increases when the intensity level is raised and decreases when it is lowered. Furthermore, the signal measured at a given point in time depends on the preceding inputs.  For example, a substantially different range of current values is measured for pulses with the same intensity level $A_3$ at $t = \SI{25}{\second}$–$\SI{30}{\second}$ ($\approx$~\SI{1.569}{\micro\ampere}) and $t = \SI{150}{\second}$–$\SI{155}{\second}$ ($\approx~1.564$–\SI{1.566}{\micro\ampere}). After corroborating that the measured photocurrent is co-determined by the present and preceding light pulses, we focus on the information the measured current at a given point in time (``Present'', pulse $k$) carries about the previous pulse (``Past'', pulse $k-1$) as defined in Fig.~\ref{fig2}(b).

For this purpose, we consider two-pulse configurations $[u(k-1), u(k)]$ within the input sequence. With three intensity levels, each “Present” pulse $u(k)$ can be preceded by any of the three “Past” pulses $u(k-1)$, giving nine possible configurations. To qualitatively demonstrate the memory effect, all photocurrent responses associated with present pulses of intensity level $A_3$ are extracted and averaged for comparison, as shown in Fig.~\ref{fig2}(c). Three distinct time-evolving current profiles emerge, each corresponding to a different past pulse, as indicated by the two-pulse configurations sketched on the left side of the plot. The configuration $[0.1,\,1.0]$ produces the lowest overall current and exhibits an initially nonlinearly increasing profile before approaching a more linear trend. In contrast, the $[1.0,\,1.0]$ configuration yields the highest current, increasing almost linearly throughout the \SI{5}{\second} window. A similar trend is observed in Fig.~\ref{fig2}(d) for present pulses of intensity level $A_2$: again, three distinct current profiles appear, with $[1.0,\,0.5]$ producing the highest values and $[0.1,\,0.5]$ the lowest. These distinct outputs for identical present-pulse intensities show that the photocurrent at pulse $k$ encodes not only the instantaneous input $u(k)$ but also the preceding input $u(k-1)$, providing the history dependence the reservoir exploits.

Having established that the photo-induced current $r(t)$ encodes information about both the present input $u(k)$ and the preceding input $u(k-1)$, we now exploit this time-evolving signal to construct a high-dimensional reservoir state. Let $t_k$ denote the onset time of the $k^{\text{th}}$ pulse and $\tau$ the time interval between two temporal readout points. As shown in Fig.~\ref{fig3}(a), the photocurrent response to the $k^{\text{th}}$ input $u(k)$ is sampled at $n$ points $r(t_k + i\tau)$ with $i = 0, \ldots, n-1$, distributed uniformly across the \SI{5}{\second} response window so as to capture the overall shape of the time-evolving photocurrent. These $n$ samples, together with a constant leading entry, form the reservoir state vector $\mathbf{x}(k) = \big(1, r(t_k), r(t_k+\tau), \ldots, r(t_k+(n-1)\tau)\big)^{\mathrm T}$, an $(n{+}1)$-dimensional vector whose leading entry supplies the bias (intercept) of the linear readout and whose remaining entries are the reservoir nodes for input $u(k)$. We use $n = 10$ reservoir nodes throughout this work, which is found to be sufficient as the recognition accuracy saturates with further increase in $n$  (see Appendix~\ref{app:nodes}).

The reservoir state vector forms the input to a linear readout that produces the present and past predictions, $y(k)$ and $y(k-1)$, through two trainable weight vectors, $\mathbf{w}_{\text{present}}$ and $\mathbf{w}_{\text{past}}$, as shown in Fig.~\ref{fig3}(b). Each prediction is the inner product of the corresponding weight vector with the state, $y(k)=\mathbf{w}_{\text{present}}\cdot\mathbf{x}(k)$ and $y(k-1)=\mathbf{w}_{\text{past}}\cdot\mathbf{x}(k)$, as illustrated in Fig.~\ref{fig3}(c); each value is then assigned to one of the three intensity levels $A_1$, $A_2$, $A_3$ by thresholding. Both predictions are read out from the same reservoir state -- only the trained weights differ -- so adapting to a new task requires retraining only the linear readout, not re-running the reservoir. The weight vectors are obtained in a single closed-form least-squares step that trains $y(k)$ and $y(k-1)$ to reproduce the present and past pulse intensity levels, with no iterative optimization (see Appendix~\ref{app:training} for the full training procedure).

Figure~\ref{fig4} presents the recognition accuracy for both the present and past input pulses as a function of the number of pulses we used to train the weights. The present pulse $u(k)$ is identified with near-perfect accuracy ($\sim 99\%$; solid green), as expected: the instantaneous photocurrent directly reflects the intensity of the present input, so this task requires no memory and serves as a reference. Recognizing the past pulse $u(k-1)$ is the nontrivial, memory-dependent task, for which we compare three configurations. Without the reservoir, when only the raw input intensities $u(k)$ are passed to the linear readout, the accuracy remains at $\sim 33\%$, which corresponds to random guessing among the three possible outputs. Using the experimentally measured photo-induced current of ErMnO$_3$ as the reservoir, the accuracy rises to $\sim 93\%$. Repeating the same training pipeline on simulated data produced by a circuit model that emulates the photocurrent response of ErMnO$_3$ (see Appendix~\ref{app:circuit} for details of the model) further raises the accuracy to $\sim 100\%$; the circuit model contains no measurement noise, which accounts for the gap to the experimental $\sim 93\%$.

The three comparisons isolate where the recognition capability originates: the reservoir transformation, not the raw input, supplies the information about $u(k-1)$, and the simulated $\sim 100\%$ accuracy shows that the photocurrent waveform already encodes enough to fully resolve the three past-pulse classes. The $\sim 7\%$ shortfall of the experimental reservoir can be attributed to measurement non-idealities, for example, readout current noise and baseline-drift artifacts, rather than to any intrinsic limitation of the reservoir material. Collectively, these results highlight the potential of photo-induced currents in ErMnO$_3$ as a physical reservoir, providing the nonlinear and memory-dependent dynamics required for reservoir computing. With improved measurement and signal-processing strategies, the experimental accuracy is expected to approach the performance predicted by simulations.

\section{Conclusion and Outlook}

In conclusion, we have demonstrated that the photo-induced current response of ErMnO$_3$ can be utilized to perform reservoir computing, enabling the recognition of past inputs within a random light-pulse sequence, a task that is inaccessible without the reservoir, since the instantaneous input carries no information about the preceding pulse. Based on our approach, a recognition accuracy of ``past events'' (here, previous light pulses) of about $93\%$ can be reached. This capability arises from the time-evolving current and relaxation dynamics, which reflect the intrinsically rich behavior of the photo-induced current in ErMnO$_3$. Looking ahead, the relaxation timescale that underlies this short-term memory can be tuned through engineering and defect control of the metal-semiconductor contacts and/or by doping the reservoir material ErMnO$_3$. Such tunability is a central asset for reservoir computing: the response time can be matched to the characteristic timescale of a given task, and several distinct timescales can be combined within a single device to increase the dimensionality of the reservoir. More broadly, our results highlight the potential of photo-induced current-based approaches for physical reservoir computing in (ferroelectric) semiconductors. Additional yet-to-be explored opportunities arise from the ferroelectric domain and domain wall structure of ferroelectrics. On the one hand, these systems give multi-facet possibilities for, e.g., optical, strain- and electric-field-driven excitations; on the other hand, the different responses of different domain states, as well as the domain walls, may be leveraged for downscaling reservoirs or to further enhance the complexity of the response.

\begin{acknowledgments}
Y.M.C.  and  D.M.  acknowledge  funding  from  the  European  Union’s  Horizon  Europe  Programme  Horizon  under the Marie Skłodowska-Curie Actions (MSCA), Grant agreement No.  101119608 (TOPOCOM). I.H. and D.M. acknowledge funding from the European
Research Council (ERC) under the European Union’s Horizon 2020 Research and Innovation Program (Grant Agreement No. 863691). M.Z. acknowledges funding from the Studienstiftung des Deutschen Volkes via a doctoral grant and the Free State of Bavaria via a Marianne-Plehn scholarship. A.~M.~ acknowledges support from the Joachim Herz Foundation through the Add-on Fellowship for Interdisciplinary Science and Transfer. K.~E.~S. acknowledges funding from the German Research Foundation (DFG) Project-ID 278162697-SFB 1242 (project B10). 
\end{acknowledgments}

\section*{Data Availability Statement}
The data that support the findings of this article are openly available~\cite{dataset_zenodo}.

\clearpage
\section*{Appendix}
\appendix

\FloatBarrier
\section{Background subtraction for photocurrent baseline-drift removal}
\label{app:baseline}

In addition to the pulse-driven response, the measured photocurrent exhibits a slow, monotonic drift of its baseline over the full acquisition, which for the pulse sequence used here lasts about \SI{33}{\minute} (400 pulses of $\sim\SI{5}{\second}$ each). The drift is small from one pulse to the next but accumulates over the run, amounting to a few percent of the mean photocurrent. Crucially, the recognition task targets only the immediate past pulse $u(k-1)$, a short-term-memory effect confined to the relaxation following a single pulse; the relevant dynamics therefore live on the few-second timescale of the inter-pulse interval, whereas the baseline drift evolves over the entire multi-minute acquisition. This slow drift carries no information about the immediate pulse history, yet, because the reservoir node states are obtained by sampling the photocurrent at fixed times within each \SI{5}{\second} pulse window (see main text), an uncorrected drift would bias these node values and effectively encode the absolute acquisition time into the readout rather than the immediate past pulse we wish to resolve. We therefore remove the drift before extracting any reservoir features, isolating the short-term dynamics relevant to the immediate past pulse.

\begin{center}
    \includegraphics{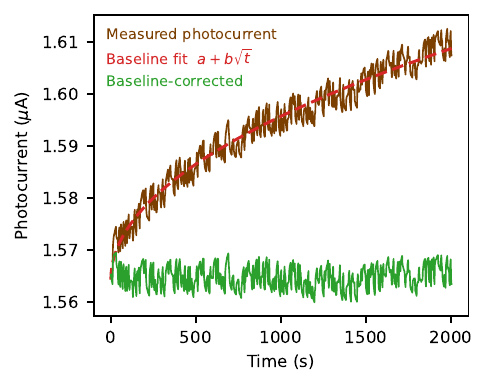}
    \captionof{figure}{Removal of the photocurrent baseline drift by background subtraction. The measured photocurrent over the active pulse window (gold) rises slowly across the $\sim\SI{33}{\minute}$ acquisition and is well described by the fitted square-root baseline $a+b\sqrt{t}$ of Eq.~\eqref{eq:supp:baseline} (red dashed). Subtracting this baseline as in Eq.~\eqref{eq:supp:corrected} yields the baseline-corrected photocurrent $I_\text{corr}$ (green): the slow drift is removed while the fast pulse-to-pulse modulation that carries the reservoir information is preserved.}
    \label{fig:supp:baseline}
\end{center}

We find that, over the active window, the baseline is well described empirically by a square-root law,
\begin{equation}
    I_\text{bg}(t) = a + b\sqrt{t - t_0},
    \label{eq:supp:baseline}
\end{equation}
where $t_0$ is the start of the window and the coefficients $a$ and $b$ are obtained for each dataset by an ordinary least-squares fit. The background-subtracted signal used for all subsequent analysis is
\begin{equation}
    I_\text{corr}(t) = I(t) - I_\text{bg}(t) + I(t_0),
    \label{eq:supp:corrected}
\end{equation}
in which the constant offset $I(t_0)$ re-anchors the corrected trace to the initial current so that the absolute current scale is preserved. As shown in Fig.~\ref{fig:supp:baseline}, the long-term drift is flattened while the pulse-to-pulse modulation that carries the reservoir information is left intact. All reservoir node states reported in the main text are extracted from $I_\text{corr}$.

\section{Training procedure for the readout layer}
\label{app:training}
The complete dataset of 400 pulses was randomly divided into a training set containing $N$ samples and a test set containing the remaining $400-N$ samples. To account for variations arising from the train–test partition, this random split was repeated independently for 1000 realizations, and the reported recognition accuracy was averaged over all realizations. For the past-pulse recognition task, only 399 pulses were available, because the first pulse has no preceding pulse and was therefore excluded.

Let the intensity of the $k^\textrm{th}$ input pulse be denoted by $u(k)\in\{A_1,A_2,A_3\}$, where $A_1=0.1$, $A_2=0.5$, and $A_3=1.0$. The three intensity levels are assigned the ordinal class labels $0$, $1$, and $2$, respectively. The class label corresponding to $u(k)$ is denoted by $c(k)$. Let $t_k$ denote the onset time of the $k^\textrm{th}$ pulse, and let $\tau$ denote the time interval between two temporal readout points. The corresponding photocurrent readouts are $r(t_k+i\tau)$ where $i=0,\ldots,n-1$. The sampled photocurrent components are standardized using the mean and standard deviation calculated from the training data. Following that the reservoir state corresponding to the $k^\textrm{th}$ input is defined as
\begin{equation*}
\mathbf{x}(k) = \big[\,1,\; r(t_k),\; r(t_k+\tau),\; \ldots,\; r(t_k+(n-1)\tau)\,\big]^{\mathrm T},
\end{equation*}
where the leading constant unity accounts for the intercept of the linear readout.

For the present pulse recognition task, the reservoir state $\mathbf{x}(k)$ is paired with the target class $c(k)$. For $N$ training samples (with pulses $k_1, k_2, \ldots, k_N$), the reservoir states and the corresponding target classes are defined as
$\mathbf{X}_{\mathrm{tr}} = \big[\mathbf{x}(k_1), \mathbf{x}(k_2), \ldots, \mathbf{x}(k_N) \big]$ and $\mathbf{c}_{\mathrm{tr}} = \big[c(k_1), c(k_2), \ldots, c(k_N) \big]$.


The linear readout weights $\mathbf{w}_{\text{present}}$ are
obtained by ordinary least-squares regression $\mathbf{w}_{\text{present}} = \underset{\mathbf{w}}{\operatorname{arg\,min}}\,
\left\|\mathbf{w} \mathbf{X}_{\mathrm{tr}} -\mathbf{c}_{\mathrm{tr}}\right\|_2^2$, computed with the \texttt{LinearRegression} class from scikit-learn.

At inference, the continuous readout output $y$ of a test state is converted into a predicted class label by thresholding, $\hat{c} = Q(y)$, with
\begin{equation}
Q(y)=
\begin{cases}
0, & y<0.5,\\
1, & 0.5\leq y<1.5,\\
2, & y\geq 1.5.
\end{cases}
\label{eq:thresholding}
\end{equation}

Recognition accuracy was evaluated on the testing as the fraction of samples for which the predicted class $\hat{c}(k)$ agrees with the corresponding target class $c(k)$. The standardization was implemented using \texttt{StandardScaler} class from scikit-learn. Recognition of the past pulse was performed analogously, using $c(k-1)$ rather than $c(k)$ as the target.

\FloatBarrier
\section{Saturation of accuracy with increasing number of reservoir nodes}
\label{app:nodes}
The dependence of the recognition accuracy on the number of reservoir nodes $n$ is shown in Fig.~\ref{fig:supp:nodes}: the past-pulse accuracy rises steeply and saturates by about $n=8$, so a larger $n$ brings no further improvement and the value $n=10$ used throughout lies on the plateau.

\begin{center}
    \includegraphics{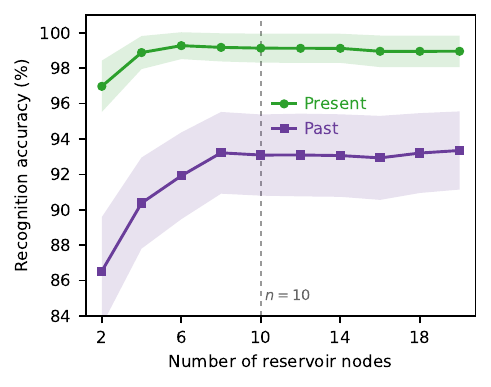}
    \captionof{figure}{Recognition accuracy as a function of the number of reservoir nodes $n$, for the uniformly distributed node sampling used throughout. The present-pulse accuracy (green) reaches $\sim 99\%$ within a few nodes, while the past-pulse accuracy (purple) rises from $n=2$ and saturates by about $n=8$; the dashed line marks the value $n=10$ used throughout, which already lies on the plateau. Curves are means over 1000 random train--test partitions; shaded bands show $\pm 1$ standard deviation.}
    \label{fig:supp:nodes}
\end{center}

\FloatBarrier
\section{Equivalent circuit model for dynamic electric response of \texorpdfstring{\ce{ErMnO3}}{ErMnO3}}
\label{app:circuit}

\begin{figure*}
    \begin{tikzpicture}[line width=1.3pt,
                    circuitikz/bipoles/thickness=1.0,
                    european]

    \node[anchor=center] at (0.0, 1.5) {\textbf{(a)}};
    \node[anchor=center] at (0.0, -2.5) {\textbf{(b)}};
    \node[anchor=center] at (8.5, 1.5) {\textbf{(c)}};

    \draw (0,0) to[short, o-*] (0.3, 0)
    to[short, -] (0.3, 1.0)
    to[R, l={$\sigma_\text{contact.}$}] (2.3, 1.0)
    to[short, -*] (2.3, 0.0)
    to[short, -*] (2.6, 0.0)
    to[short, -] (2.6, 1.0)
    to[R, l={$\sigma_\text{DW}$}] (4.6, 1.0)
    to[short, -*] (4.6, 0.0)
    to[short, -*] (4.9, 0.0)
    to[short, -] (4.9, 1.0)
    to[R, l={$\sigma_\text{UDR}$}] (6.9, 1.0)
    to[short, -*] (6.9, 0.0)
    to[short, -o] (7.2, 0.0);
    \draw (4.9, 0.0)
    to[R, l={$\sigma_\text{dc}$}] (6.9, 0.0);
    \draw (0.3, 0.0)
    to[short, -] (0.3, -1.0)
    to[C, l_={$\varepsilon_\text{contact}$}] (2.3, -1.0)
    to[short, -] (2.3, 0.0);
    \draw (2.6, 0.0)
    to[short, -] (2.6, -1.0)
    to[C, l_={$\varepsilon_\text{DW}$}] (4.6, -1.0)
    to[short, -] (4.6, 0.0);
    \draw (4.9, 0.0)
    to[short, -] (4.9, -1.0)
    to[C, l_={$\varepsilon_\text{bulk}$}] (6.9, -1.0)
    to[short, -] (6.9, 0.0);

    \draw (0.0, -3.0)
    to[short, o-*] (1.0, -3.0)
    to[short, -] (1.0, -2.5)
    to[R, l_=$R_1$] (3.0, -2.5)
    to[short, -*] (4.0, -2.5)
    to[short, -] (4.0, -3.5)
    to[C, l^=$C_1$] (4.0, -5.5);
    \draw (0.0, -5.5)
    to[short, o-*] (3.0, -5.5)
    to[short, -*] (4.0, -5.5)
    to[short, -o] (7.2, -5.5);
    \draw (1.0, -3.0)
    to[short, -] (1.0, -3.5)
    to[R, l_=$R_2$] (3.0, -3.5)
    to[C, l_=$C_2$] (3.0, -5.5);
    \draw (4.0, -2.5)
    to[short, -] (4.2, -2.5)
    to[R, l_=$R_3$] (6.2, -2.5)
    to[short, -*] (6.2, -3.0)
    to[short, -o] (7.2, -3.0);
    \draw (3.0, -3.5)
    to[short, *-] (4.2, -3.5)
    to[R, l_=$R_4$] (6.2, -3.5)
    to[short, -] (6.2, -3.0);

    \node[anchor=center] at (0.0, -4.3) {$V_\text{in}(t)$};
    \node[anchor=center] at (7.2, -4.3) {$V_\text{out}(t)$};

    \begin{axis}[
        at={(9.5cm,-5.5cm)},
        height=8.1cm,
        xlabel={Time (s)},
        ylabel={Current (pA)},
        clip mode=individual,
    ]
        \addplot[
            only marks,
            mark=o,
            draw=black,
            restrict expr to domain={rawx}{-1:1.6},
        ] table [
            x index=1,
            y index=2,
            col sep=comma
        ]{data/decay_fit_visualization.csv};
        \addplot[
            color=red,
            restrict expr to domain={rawx}{0.01:1.6},
            thick,
        ] table [
            x index=1,
            y index=3,
            col sep=comma
]			{data/decay_fit_visualization.csv};
    \end{axis}
\end{tikzpicture}
    \caption{Equivalent circuit modeling of the dynamic electric response of \ce{ErMnO3}. (a) Model proposed by Ref.~\onlinecite{puntigam_insulating_2021} for \ce{ErMnO3} bulk material contacted by two metal electrodes. (b) Modified circuit model applied in the current paper. Changes are explained and justified in the text below. (c) Photocurrent between two metal electrodes (open black dots) after switching off the illuminating white light source at $t = 0$. Drawn-through red line is the double-exponential fit used to obtain the circuit element values.}
    \label{fig:supp:circuits}
\end{figure*}
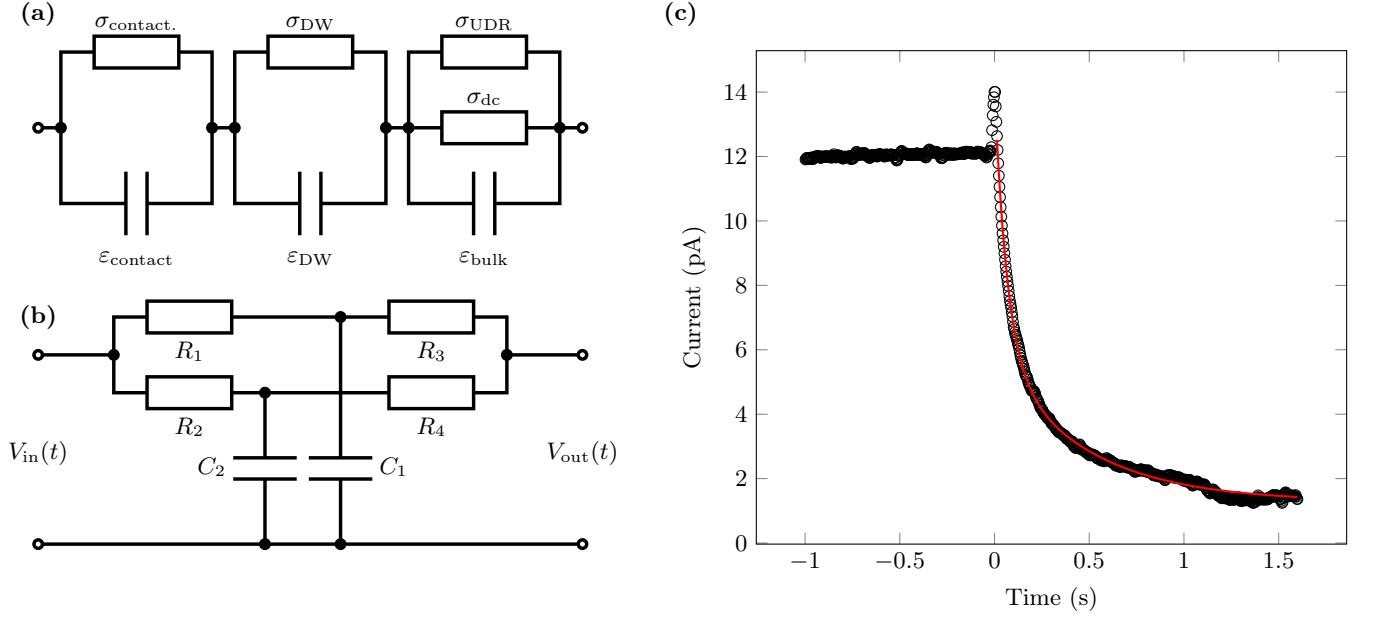

The dynamic electric response of \ce{ErMnO3} has been investigated in Ref.~\onlinecite{puntigam_insulating_2021} by dielectric spectroscopy and the equivalent circuit model shown in Fig.~\ref{fig:supp:circuits}(a) has been concluded. Three contributions are identified, each represented by a separate RC-circuit:

\begin{enumerate}
    \item the bulk material. While the intrinsic dielectric constant $\varepsilon_\text{bulk}$ creates a capacitive contribution, the real part resistance consists of a frequency independent part $\sigma_\text{dc}$ and a so-called universal dielectric response (UDR) \onlinecite{jonscher_universal_1977} with power-law scaling with respect to frequency $\omega$, $\sigma_\text{UDR} \propto \omega^\nu$, $0 < \nu < 1$.
    \item the contacts between electrodes and bulk material, generating via a Maxwell-Wagner relaxation a RC-circuit contribution characterized by $\sigma_\text{contact}$ and $\varepsilon_\text{contact}$.
    \item the ferroelectric domain walls, acting as additional capacitive barrier layers, characterized by $\sigma_\text{DW}$ and $\varepsilon_\text{DW}$.
\end{enumerate}

Based on these considerations, we derive a modified circuit representing the photocurrent scenario. It is developed by intention pictorially and can be abstracted and in principle treated more rigorously by signal processing methods:

\begin{itemize}
    \item In the first approximation, the photocurrent in many photovoltaic materials is proportional to the photon flux. The photovoltaic effect is therefore modeled as a linear process and the input signal is represented by an ideal voltage source with $A_1 = \SI{0.1}{V}$, $A_2 = \SI{0.5}{V}$ and $A_3 = \SI{1.0}{V}$.
    \item Since linear scaling of the signals doesn't play a role for the modeling as it is covered by the linear readout layer, the most important property of the RC circuits is their time constant. Experimental results of Ref.~\cite{puntigam_insulating_2021} confirm that they span over several orders of magnitude in time. Since we investigate processes on the time scale of the slowest relaxation, the contribution of the fastest contribution can be neglected, and we restrict our model to two RC-circuits.
    \item In the following step, the model parameters will be determined from experimental data on the transient response after an abrupt change of light intensity. To simplify the fitting process, we arrange the RC-circuits in parallel compared to the series circuit in the original case, introducing no relevant deviations within the frequency range of interest.
\end{itemize}

Finally, we end up with the circuit shown in Fig.~\ref{fig:supp:circuits}(b), leaving the task to determine the model parameters. As previously mentioned, the absolute signal values do not influence the outcome since the linear output layer can cover arbitrary linear rescaling. This introduces the freedom to choose some values arbitrarily and the remaining parameters are determined based on the transient response shown in Fig.~\ref{fig:supp:circuits}(c). After switching off the photon flux, the current is described by a double exponential decay parametrized as:

\begin{equation}
    I(t) = I_0 + I_1 \exp(-t / \tau_1) + I_2 \exp(-t / \tau_2).
\end{equation}

In the present example, the following values are obtained: $I_0 = \SI{1.279 \pm 0.002}{pA}$, $I_1 = \SI{8.287 \pm 0.013}{pA}$, {$I_2 = \SI{4.567 \pm 0.012}{pA}$}, $\tau_1 = \SI{65.90 \pm 0.20}{ms}$ and $\tau_2 = \SI{467.9 \pm 1.7}{ms}$.

After choosing some parameters only affecting the absolute response values,

\begin{equation*}
    R_1 = R_2 = \SI{3e5}{\ohm}, \ R_3 = \SI{3e7}{\ohm}
\end{equation*}

the remaining properties can be calculated from those and the fit parameters:

\begin{equation*}
    C_1 = \tau_1/R_1, \ C_2 = \tau_2/R_2, \ R_4 = R_3 A_2/A_1.
\end{equation*}

For the comparison to experimental results, the same pulse series as utilized in the experiment has been applied on the input side of the circuit shown in Fig.~\ref{fig:supp:circuits}(b) and corresponding output is obtained by enforcing Kirchhoff's current law on a series of discrete time steps.

\bibliography{references}

\end{document}